\documentstyle[preprint,aps]{revtex}
\begin{document}
\draft
\preprint{}
\title{The field theory of Skyrme lattices 
in quantum Hall ferromagnets}
\author{M. Abolfath$^{1,2,3}$, and M. R. Ejtehadi$^3$}
\address{$^1$Department of Physics, Indiana University, Bloomington, 
Indiana 47405}
\address{$^2$Center for Theoretical Physics and Mathematics, 
P.O.Box 11365-8486, Tehran, Iran}
\address{$^3$
Institute for Studies in Theoretical Physics and Mathematics
P.O. Box 19395-1795, Tehran, Iran
%Department of Physics, Sharif University of Technology,
%Tehran, P.O.Box 11365-9161, Tehran, Iran
}
\date{\today}
\maketitle

\begin{abstract}
We report the application of the nonlinear $\sigma$ model 
to study the multi-skyrmion problem in the
quantum Hall ferromagnet system. We show that the ground
state of the system can be described by a 
ferromagnet triangular Skyrme lattice
near $\nu=1$ where skyrmions are extremely dilute. 
We find a transition into 
antiferromagnet square lattice by increasing the 
skyrmion density and therefore $|\nu-1|$.
We investigate the possibility that the square Skyrme lattice deforms 
to a single skyrmion with the same topological charge
when the Zeeman energy is extremely smaller than the Coulomb energy.
We explicitly show that the energy of a skyrmion with charge two is less than the 
energy of two skyrmions each with charge one when 
$g \leq g_c$. By taking the quantum fluctuations into account, 
we also argue the possibility of the existence of a %SMG
non-zero temperature
Kosterlitz-Thouless and a superconductor-insulator phase transition. 
\end{abstract}
%%%%%%%%%%%%%%%%%%%%%%%%%%%%%%%%%%%%%%%%%%%%%%%%%%
\section{Introduction}
Recently, there have been a number of experimental 
\cite{barrett,jpe-skyrme,optical1,bayot}
and theoretical \cite{Sondhi,Moon,Yang,Abolfath} 
investigations of the skyrmions in the integer and 
fractional quantum Hall effect (QHE) in 
2D electron gas (2DEG). It has been shown \cite{Brey,Green,Shankar,Cote97}
that the ground state of a two-dimensional electron system at
the Landau level filling factor near $\nu=1$ 
is a Skyrme (soliton) lattice.
%The Skyrme lattice in quantum Hall ferromagnet has been studied by 
%using the random phase approximation \cite{Brey}.
%In this letter we implicate the classical field theory
%to explain the existence of the Skyrme lattice.
Following Sondhi {\em et al.} \cite{Sondhi} and Moon {\em et al.} 
\cite{Moon}, one may show that
the quantum Hall ferromagnet can be described by an effective
nonlinear $\sigma$ (NL$\sigma$) model.
In this article we also make use of the effective field theory of the
quantum Hall NL$\sigma$ model \cite{Sondhi,Moon,Abolfath,Brey,Green}
to study a multi-skyrmion %SMG
 quantum Hall system.
Our results for a multi-skyrmion system show %SMG s
 that the ground state has %SMG a 
long range order. We take the advantage of particle-hole symmetry
\cite{Fertig94} (which is equivalent %SMG
to skyrmion-antiskyrmion duality 
in classical field theory) 
to study the skyrmions associated with holes ($\nu \leq 1$).
We will show that within $\nu_{1c} < \nu \sim 1$, 
the ground state is a triangular lattice  
in order to reduce the Coulomb energy since,
the Zeeman energy is negligible. Increasing the skyrmion % SMG s
 density 
in such a way that $0 << \nu_{2c} < \nu < \nu_{1c}$, 
leads to the formation of %SMG forming 
a square lattice. 
The square lattice is favored by the gradient and Zeeman energy 
in order to allow %SMG make
 the antiferromagnetic 
alignment of the skyrmion's orientation via the attractive XY 
interaction between them. 
This can be anticipated by the frustration of the
skyrmionic hedgehog fields within the triangular 
lattice which is the %SMG
favorable orientation for the Coulomb energy.
This leads to a structural phase transition due to 
varying %SMG
the Landau level filling factor.
By taking the quantum fluctuations into account, 
we also argue the possibility of the existence of a %SMG
non-zero temperature
Kosterlitz-Thouless and a superconductor-insulator phase transition. 
We show that the 2DES with the typical Zeeman energy at filling factor
smaller than one (the validity range of the square lattice) 
are in the insulating phase.
The attractive XY interaction makes a tendency for collapsing the 
single skyrmions which are located in different positions. 
It favors recombination of the $N$-seperated single skyrmions 
and forming a single skyrmion with topological charge $N$.
The competition between attractive XY interaction and repulsive 
Coulomb interaction leads to the stability of the multi-skyrmion system
in some circumstances.
More explicitly, we investigate this for a special two single skyrmions 
system numerically.
We demonstrate that in a special two single-skyrmion case at
extremely small values of the Zeeman energy, %SMG the 
(depends on the filling factor) collapse can %SMG be 
occur through %SMG hence 
the formation of a single skyrmion
with the topological charge %SMG
two.

%%%%%%%%%%%%%%%%%%%%%%%%%%%%%%%%%%%%%%%%%%%%%%%%%%%%%%%%%%%%%%%%%%
\section{An overview on NL$\sigma$-model in QHF}
In the NL$\sigma$-model approach to %SMG of 
QHF, the spin of an electron may
be described classically by a unit vector whose %SMG
 direction may be
changed continuously %SMG
in the space. In this representation, the effective
Hamiltonian is functional of the unit vector ${\bf m}({\bf r})$
\cite{Sondhi,Moon}
\begin{equation}
E[{\bf m}]=E_0[{\bf m}]+E_z[{\bf m}]+E_c[{\bf m}] \; ,
\label{eq1}
\end{equation}
where $E_0[{\bf m}]$ and $E_z[{\bf m}]$ are the conventional NL$\sigma$ model 
and the Zeeman energy respectively. 
The last term $E_c[{\bf m}]$ is the Coulomb energy 
due to the connection between excess electric charge density
and skyrmion topological density\cite{Sondhi,Moon}
\begin{mathletters}
\label{eq2}
\begin{equation}
E_0[{\bf m}]=\frac {\rho_s}{2}\; \int d^{2}r\, (\nabla{\bf m})^2\; , 
\label{eq2a}
\end{equation}
\begin{equation}
E_z[{\bf m}]=\frac{\tilde{g}}{2 \pi \ell_0^2} 
\int d^{2}r\, [ \; 1- m_z({\bf r}) \; ]\; , 
\label{eq2b}
\end{equation}
\begin{equation}
E_c[{\bf m}]={e^2\over 2 \epsilon} \int d^{2}r \int d^{2}r^\prime \,
\frac{\rho({\bf r})\rho({\bf r}^\prime)}{|{\bf r}-{\bf r}^\prime|} \; .
\label{eq2c}
\end{equation}
\end{mathletters}
Here $\rho_s=e^2/(16\sqrt{2\pi}\epsilon\ell_0)$ is the spin stiffness
(assuming zero layer thickness for the 2DEG) which arises from the Coulomb
exchange energy,
$\epsilon$ is the background dielectric constant of the semiconductor,
$\tilde{g}=g e^2/(2\epsilon\ell_0)$ is the Zeeman term, $g$ is the
effective gyromagnetic ratio, and $\ell_0$ is the magnetic length. 
The charge density is given by \cite{Sondhi,Moon}
$\rho({\bf r})=\frac{-\nu}{8 \pi} \; \epsilon_{\alpha \beta} \; 
{\bf m({\bf r})} 
\cdot [\partial_\alpha{\bf m({\bf r})} \times 
\partial_\beta{\bf m({\bf r})}]$ 
which is equal to the filling factor times
the topological $O(3)$ spin texture density of the 
quantum Hall ferromagnet (QHF). 
The total skyrmion charge denoted by $Q$
is an integer-valued topological invariant. It can be determined
by the integration upon the charge density and classified by homotopy group
of a 2D-sphere respectively.

The lowest energy skyrmion solution, ${\bf \tilde{m}}({\bf r})$,
has to satisfy a non-linear differential equation which can be obtained
by minimizing the energy in Eq.(\ref{eq1}) with respect to ${\bf m}$
\begin{equation}
\rho_s \Bigl( -\nabla^2  + {\bf \tilde{m}} \cdot 
\nabla^2 {\bf \tilde{m}} \Bigl){\rm \tilde{m}}_\mu 
-\frac{\tilde{g}}{2\pi \ell_0^2} (\delta_{z\mu}-
{\rm \tilde{m}}_z{\rm \tilde{m}}_\mu)
-\frac{\nu}{4\pi} \epsilon_{\alpha \beta} \{ \partial_\alpha V({\bf r}) \}
({\bf \tilde{m}} \times \partial_\beta {\bf \tilde{m}})_\mu=0 \; ,
\label{eq3}
\end{equation}
where $V({\bf r})$ is Hartree potential (the exchange potential resides in
$\rho_s$)
\begin{equation}
V({\bf r})=\frac{e^2}{\epsilon_0}
\int d{\bf r}^\prime \frac{\tilde{\rho}({\bf r}^\prime)}
{|{\bf r}-{\bf r}^\prime|} \; ,
\label{eq4}
\end{equation}
and $\tilde{\rho}$ is the skyrmion charge density associated with the 
minimum energy solution, ${\bf \tilde{m}}({\bf r})$.
The solutions of Eq.(\ref{eq3}) can be classified by the skyrmion charge
$Q=\int d{\bf r} \; \rho({\bf r})$. From now 
and for the sake of simplicity, we remove 
the tilde over the classical solution and denote it by ${\bf m}$.
It is easy to find the following equation of motion of the optimal skyrmion
by making use of cross product of ${\bf m}$ upon Eq.(\ref{eq3})
\begin{equation}
\partial_\alpha J_\alpha^\lambda = \frac{\tilde{g}}{2\pi\ell_0^2}
(\hat{z}\times{\bf m})_\lambda,
\label{eq5}
\end{equation}
where
\begin{equation}
J_\alpha^\lambda=\rho_s({\bf m}\times\partial_\alpha{\bf m})_\lambda
-\frac{\nu}{4\pi} V({\bf r}) 
\epsilon_{\alpha \beta}\partial_\beta m_\lambda.
\label{eq6}
\end{equation}
One may immediately read off from Eq.(\ref{eq5}) that 
$\partial_\alpha J_\alpha^3 = 0$.
%, i.e. a conservation law.
We may define the ground state of QHF at 
$\nu=1$ as a vacuum of skyrmionic spin textures %SMG
 where all spins are aligned
along the magnetic field direction, i.e. the $\hat{z}$-axis.  %SMG
Note that in the
absence of the Zeeman energy, the alignment of spins along an arbitrary
axis occurs due to spontaneous %SMG
global $O(3)$ symmetry breaking \cite{Moon}
hence the minimizing the electrons exchange Coulomb energy.
This is also the state of spins far from the center of the skyrmions.
Therefore the number of %SMG the 
skyrmions (antiskyrmions) is %SMG
counted by $|N-N_\phi|$.

Before we consider a lattice of skyrmions we 
need to have the correct shape of a single skyrmion. 
A single skyrmion is a topological optimal solution of 
Eq.(\ref{eq4}) with $Q=1$.
For our purposes, it is convenient %SMG
 to
parameterize the unit vector ${\bf m}$ by
\begin{equation}
{\bf m} = (\varphi_x, \varphi_y, \sqrt{1-\overline{\psi}\psi})
\label{eq6.1}
\end{equation}
where $\psi = \varphi_x + i \varphi_y$. 
Near the core of the skyrmion we do not expect that the shape of
the skyrmion is influenced much by the magnetic field. However,
for large distances from the core, the Zeeman energy becomes
dominant and we need to consider its effect where
the direction of the unit vector ${\bf m}$ 
is close to the vacuum, namely, $\hat{z}$-axis. 
Taking the limit of small $\psi$, we can expand the ${\bf m}$ up to 
quadratic order in the $\psi$. Putting this in Eq.(\ref{eq6.1}) gives
\begin{equation}
E[\psi] = \int d{\bf r} \left(\frac{-\rho_s}{2} \overline{\psi}
\nabla^2 \psi + \frac{\tilde{g}}{4\pi\ell_0^2} \overline{\psi} \psi\right),
\label{eq6.2}
\end{equation}
which leads to the equation of motion

\begin{equation}
-\rho_s\nabla^2 \psi
+ \frac{\tilde{g}}{2\pi\ell_0^2} \psi = 0.
\label{eq6.21}
\end{equation}
Introducing $\kappa^2 = \tilde{g}/(2\pi\ell_0^2\rho_s)$
we simply have the equation $-\nabla^2 \psi + \kappa^2 \psi=0$.
We are interested in the `vortex' solution 
$\psi = 2\partial_{z} \xi$, 
where $\partial_{z} = (\partial_x + i \partial_y)/2$ 
and $z=x+iy$. (If $\tilde{g}=0$, this would result in $\nabla^2 \xi=0$ with
the solution $\xi \propto \ln(r)$ and therefore $\psi \propto z/r^2$).
Substituting this we find $-\nabla^2 \xi + \kappa^2 \xi =0$ or
$\xi \propto e^{-\kappa r}/\sqrt{r}$ and therefore that 
$\psi \propto z e^{-\kappa r}/r^{3/2}$ for $r\rightarrow\infty$.
The most important of this part is of course the exponential
(as opposed to, algebraic if $\tilde{g}=0$), fall-off \cite{Henk}. 
The dynamics of a skyrmion spin texture in NL$\sigma$-model 
may be incorporated via the Wess-Zumino action \cite{fradkin}.
The result of expansion for the single skyrmion's Wess-Zumino term is
\begin{equation}
S_{WZ} = \frac{\hbar}{4\pi\ell_0^2} \int d\tau
\int dt \; {\bf m}\cdot(\partial_\tau{\bf m}\times\partial_t{\bf m})
= \frac{\hbar}{8\pi\ell_0^2} \int dt \int d{\bf r} \;
\overline{\psi}({\bf r}, t) i \frac{\partial}{\partial t} \psi({\bf r}, t),
\label{12.1}
\end{equation}
where we keep the quadratic terms in Eq.(\ref{12.1}). 
At this level of approximation,
the effective action may be obtained by Eq.(\ref{eq6.2}) and Eq.(\ref{12.1})
where the Lagrangian density is 
\begin{equation}
{\cal L} = \overline{\psi}({\bf r}, t) \left(\frac{\hbar}{8\pi\ell_0^2} 
i \frac{\partial}{\partial t} +
\frac{\rho_s}{2} \nabla^2 - \frac{\tilde{g}}{4\pi\ell_0^2} \right)
\psi({\bf r}, t),
\label{eq6.3}
\end{equation}
and $S = \int dt \int d{\bf r} {\cal L}$.
The optimal solution of $\psi$ is then identical to the solution of the %SMG
time dependent Schr\"odinger equation where the external potential is 
proportional to the Zeeman splitting factor. Then the single skyrmion behaves
approximately like a quantum mechanical point particle far from its core. 

%%%%%%%%%%%%%%%%%%%%%%%%%%%%%%%%%%%%%%%%%%%%%%%%%%%%%%%%%%%%%%%%%%%%%%
\section{Skyrmion interaction}

We can now consider the interaction between skyrmions by generalizing 
our linearized energy functional. For that one should note
that for the optimal single skyrmion spin texture in Eq.\ (\ref{eq3}) 
we may choose a particular orientation. 
In general the texture can be rotated without costing any %SMG the
energy. Since the system shows the $U(1)$ symmetry, any valid 
skyrmion spin texture can be obtained %SMG emerged via 
by rotating all spins about
the $\hat{z}$-axis by angle $\chi$. Therefore the state,
${\bf m}' = \exp(i \chi \hat{I}_z){\bf m}$, is also an optimal solution
of the Hamiltonian then $\hat{I}_z \equiv -i \partial/\partial\chi$ is the 
generator of the rotation along the $\hat{z}$-axis in the internal space. 
One may expect that it contributes to the Hamiltonian through 
$(1/2\Lambda_0)(-i\partial/\partial\chi - \xi_0)^2$ where $\Lambda_0$
is the moment of inertia %SMG
of the single skyrmion. 
This is the leading term of the total energy which is expanded 
with respect to the number of reversed spin, $\xi$.
In general the dimensional analysis shows $E_z \propto \xi$ and 
$E_c \propto 1/\sqrt{\xi}$.
The optimal value of the number of reversed spin, 
$\xi_0 = E_z/(2\tilde{g})$, can be evaluated by minimization of the
total energy with respect to $\xi$
\begin{equation}
\xi_0 \equiv \frac{1}{4\pi\ell_0^2} \int d{\bf r}
[1-m_z({\bf r})]
\label{xi}
\end{equation}
where $m_z({\bf r})$ is the optimal solution of Eq.(\ref{eq3}) 
\cite{Abolfath} corresponding to the given Zeeman splitting factor. 
This leads to the optimal Coulomb and Zeeman energy and then the predicition 
$E_c/E_z = 2$ hence 
$\Lambda_0 \equiv \left(d^2 E/d\xi^2\right)^{-1}_{\xi_0}
= E_z/(6\tilde{g}^2)$. 
Our goal is now to calculate the interaction
energy between skyrmions with different orientations. 
Following Piette {\em et al.} \cite{Piette}, 
we start with the conventional NL$\sigma$ 
model to find out the proper superposition rule for skyrmions. 
In the absence of the Zeeman and Coulomb energies,
the energy functional is scale invariant and one may find the optimal
solutions analytically. In this case, it can be shown that any 
analytic complex polynomial defines an optimal solution 
\cite{rajaraman,Blavin,Woo} 
\begin{equation}
{\bf m(r)}=\left(\frac{2w_x}{1+|w|^2}, \frac{2w_y}{1+|w|^2},
\frac{1-|w|^2}{1+|w|^2}\right),
\label{1}
\end{equation}
where 
$w = w_x + i w_y$ is any $Q$-sector analytical function.
One may decompose $w$ into a series of analytical functions each
with $Q=1$
\begin{equation}
w = \sum_{j=1}^N u_j.
\label{2}
\end{equation}
Any $Q=1$ analytical function represent a single-skyrmion, 
then the number of skyrmions $N$ is clearly
the total skyrmions topological charge, i.e. $N=Q$.
Eq.(\ref{2}) denotes a sequence of order parameters 
${\bf m}_j({\bf r})$ in configuration space. 
It is convenient to parameterize the ${\bf m}_j({\bf r})$ by
\begin{equation}
{\bf m}_j({\bf r}) = 
\left( \sin\eta_j({\bf r})  \cos\zeta_j({\bf r}), 
\sin\eta_j({\bf r})  \sin\zeta_j({\bf r}),
\cos\eta_j({\bf r}) \right),
\label{2.1}
\end{equation}
where $\eta_j({\bf r})$ and $\zeta_j({\bf r})$ are polar and 
azimuthal field variable associated with $j$th skyrmion.
One may define \cite{Abolfath,Piette} 
$\zeta_j = \varphi - \chi_j$ where $\varphi$ is the standard azimuthal
angle, e.g. ${\bf r}=(r\cos\varphi, r\sin\varphi)$
and $\chi_j$ is skyrmions orientation and measuring the deviation
from the standard hedgehog fields. 
If we have a gas of skyrmions far away from each other,
the total energy is invariant under variation of 
skyrmions orientation. For finite separation we expect a coupling  
between skyrmions due to different orientations.
Here we consider a situation 
where ${\bf m}_i$ are localized and well separated. 
This is valid for dilute skyrmions in a quantum Hall system, e.g. 
$\nu_{2c} < \nu < \nu_{1c}$. One may
divide the configuration space (${\rm R^2}$) into $N$ regions such
that $u_i$ is significant in region $i$ and small in others
\begin{equation}
u_{i\mu}=\frac{{\bf m}_{i\mu}}{1+\hat{z}\cdot{\bf m}_i}.  %SMG 
\label{3}
\end{equation}
Here %%SMG
$\mu=(x,y)$ and 
\begin{equation}
{\bf m}_j=({\bf \varphi}_x^j, \varphi_y^j, 
\sqrt{1-\Phi_j\cdot\Phi_j}),
\label{4}
\end{equation}
where $j \neq i$ and
$\Phi_j=(\varphi_x^j, \varphi_y^j)$
\begin{equation}
u_{j\mu}=\frac{\varphi_\mu^j}{1+\sqrt{1-\Phi_j\cdot\Phi_j}}.
\label{4.1}
\end{equation}
Substituting Eq.(\ref{2})-Eq.(\ref{4.1}) into Eq.(\ref{1}) and expanding
${\bf m}_j$ up to $\varphi_j$, we may find ${\bf m}$ in $i$th region
\begin{equation}
{\bf m}={\bf m}_i + \Omega_i\times{\bf m}_i +
\frac{1}{2}\Omega_i\times(\Omega_i\times{\bf m}_i) +  
{\cal O}(\Omega^3),
\label{5}
\end{equation}
here Eq.(\ref{5}) describes an infinitesimal rotation of ${\bf m}_i$
about $\Omega_i$ axis where
\begin{equation}
\Omega_i=\frac{1}{2}{\bf m}_i\times\{
(1+\hat{z}\cdot{\bf m}_i) \Phi_{\rm eff}^i - 
({\bf m}_i\cdot\Phi_{\rm eff}^i)\hat{z}\},
\label{6}
\end{equation}
and 
\begin{equation}
\Phi_{\rm eff}^i({\bf r}) = \sum_{j \neq i}^N \Phi_j({\bf r}).
\label{7}
\end{equation}
Therefore the effect of the other skyrmions on the specific skyrmion
is the same as a single skyrmion %SMG s
 with charge $Q-1$ via the effective
linear field, $\Phi_{\rm eff}$. 
In order to study the effect of the Zeeman and Coulomb energy,
we make the ansatz that the above superposition rule is valid 
for skyrmions far from each other even in the presence of full interaction.
This can be taken into account by Eq.(\ref{5}) and evaluating
the total energy of multi-skyrmion in QHF.
It is obvious that the total energy can be divided into energies in 
separated regions. We assume that the interaction between the skyrmions
is weak, hence $\Omega_i={\bf m}_i \times \Phi_{\rm eff}^i$.
One may obtain easily the total energy by
redoing the same calculation for all separated regions                                                 
and sum over energies
\begin{equation}
E[{\bf m}]=\sum_{i=1}^N \left(
E[{\bf m}_i] +
\int_i d{\bf r} \; \overline{\psi}_{\rm eff}^i \{ -\frac{\rho_s}{2} 
\nabla^2 + \frac{\tilde{g}}{4\pi\ell_0^2} \} \psi_{\rm eff}^i \right) 
+ V_{\rm eff}[{\bf m}],
\label{8}
\end{equation}
where $\psi_{\rm eff}^i\equiv \Phi_{x \; \rm eff}^i + 
i \Phi_{y \; \rm eff}^i$
and $E_{\rm eff}^C[{\bf m}]$ is the effective Coulomb 
interaction between skyrmions. Note that in the absence of
the Coulomb term, the saddle point solution 
associated with the scalar field, $\psi^i$, are vortices, i.e.
$-\nabla^2 \psi^i + \kappa^2 \psi^i=0$.
One may divide the total energy, Eq.(\ref{8}), into two parts,
the self energy of skyrmions and interaction energy which are
designated by $T$ and $V$ respectively
\begin{equation}
T[{\bf m}]=\sum_{i=1}^N \left(
E[{\bf m}_i] +
\int_i d{\bf r} \sum_{j \neq i}^N
\overline{\psi}^j \{-\frac{\rho_s}{2} \nabla^2
+ \frac{\tilde{g}}{4\pi\ell_0^2} \} \psi^j \right).
\label{9}
\end{equation}
The first term in Eq.(\ref{9}) is the total self energy of the isolated 
skyrmions and 
the second term is the effect of their tail in other regions, %SMG
i.e., the
contribution of their kinetic energy in regions far from the core.
Here we are interested to study the effect of 
interaction in a system of many skyrmions and their physical relevance.
One may find the effective interaction between the skyrmions by making
use of the Stokes theorem and neglecting 
the next nearest neighbor %SMG 
terms in Eq.(\ref{8})
which are described by the terms like
$\int_i d{\bf r} \sum_{j \neq i}\sum_{k \neq j} \Phi_j \cdot \Phi_k$
and $\int_i d{\bf r} \sum_{j \neq i}\sum_{k \neq j}
\partial_\alpha\Phi_j\cdot\partial_\alpha\Phi_k$ then
\begin{equation}
V_{\rm eff}[{\bf m}]= 
E_{\rm eff}^0[{\bf m}] + E_{\rm eff}^C[{\bf m}],
\label{11.0}
\end{equation}
and
\begin{equation}
E_{\rm eff}^0[{\bf m}] = \sum_{i=1}^N
\int_i d{\bf r} \; \partial_a \{J_a^{\lambda(i)} \Omega_{i\lambda} \} 
= \frac{\rho_s}{2} \sum_{<ij>} \int_i d{\bf r} \overline{\psi}^j 
\{ \nabla^2 - \kappa^2 \} \psi^i, 
\label{11}
\end{equation}
where ${\bf J}^{(i)}$ has been defined for $i$th skyrmion by Eq.(\ref{eq6}).
$E^0_{\rm eff}[{\bf m}]$ is the contribution of the gradient and Zeeman energy
to the effective interaction. It describes a system of interacting dipoles.
Expanding the charge density of skyrmions, $\rho$, in terms of $\psi$,
leads to the effective Coulomb interaction
$$E_{\rm eff}^C[{\bf m}] =  
\frac{e^2}{2\epsilon_0} \sum_{i \neq j}
\int_i d{\bf r} \int_j d{\bf r^\prime}
\frac{\rho_i({\bf r}) \rho_j({\bf r}^\prime)}{|{\bf r}-{\bf r}^\prime|}$$
\begin{equation}
+ \epsilon_{\mu\nu} \frac{\nu e^2}{4\pi\epsilon_0}
\sum_{i \neq j} \{\int_i d{\bf r} 
\partial_\mu V^j({\bf r})\partial_\nu{\bf m}_i \cdot \Omega_j
+ \int_j d{\bf r}
\partial_\mu V^i({\bf r})\partial_\nu{\bf m}_j \cdot \Omega_i
\},
\label{11.1}
\end{equation}
where $\rho_i=\frac{-\nu}{8 \pi} \; \epsilon_{\alpha \beta} \; 
{\bf m}_i \cdot [\partial_\alpha{\bf m}_i \times 
\partial_\beta{\bf m}_i]$. 
The first and second term in Eq.(\ref{11.1}) are the 
Coulomb energy due to the
monopole and dipole counterparts of skyrmions respectively. 
The former %SMG
falls off like $R^{-1}$ whereas the latter 
falls as $R^{-2}$ 
where the distance between two skyrmions is denoted by $R$.
%%%%%%%%%%%%%%%%%%%%%%%%%%%%
%One may show that the dipole Coulomb term contributes 
%via the quartic interaction. Then we may neglect it in the 
%gaussian level of approximation.
%
% //RAMIN: IS THIS TRUE?//
%%%%%%%%%%%%%%%%%%%%%%%%%%%%%%%%%%%%%%%%%%%%%
Unlike the dense skyrmions in which the effect of the dipole term is crucial,
the monopole term dominates for an extremely dilute (i.e. $\nu \simeq 1$)
skyrmions system. Neglecting the Coulombic dipole term, 
one may evaluate the integrals in Eq.(\ref{11}) by 
applying the techniques that %SMG
were developed for a pair of skyrmions
by Piette {\em et al.} \cite{Piette}
\begin{equation}
V_{\rm eff}[{\bf m}] = 
\frac{e^2}{2\epsilon_0} \sum_{i \neq j}
\int_i d{\bf r} \int_j d{\bf r^\prime}
\frac{\rho_i({\bf r}) \rho_j({\bf r}^\prime)}{|{\bf r}-{\bf r}^\prime|}
+ \frac{c^2\tilde{g}}{4\pi^2} \sum_{<ij>}
\cos(\chi_j-\chi_i) K_0(\kappa |{\bf R}_j-{\bf R}_i|),
\label{12}
\end{equation}
where $\chi_j-\chi_i$ and $R_j-R_i$ 
describe the in-plane relative orientation and effective distance 
between skyrmion $j$th and $i$th. The first term in Eq.(\ref{12}) is 
the electrostatic monopole term independent of the relative orientation.
The normalization factor, $c=2\sqrt{2}$, is calculated for a single skyrmion.
%by Piette {\em et al.} \cite{Piette}.
The second term in Eq.(\ref{12}) describes a classical XY-model 
where the minimum energy configuration specifies the 
relative orientation corresponds to $\chi_j-\chi_i=\pi$. 
$K_0(x)$ is the modified Bessel function, hence 
the coupling between the site $i$ and $j$ decays exponentially. 
%\null From %SMG CAN NOT START A LINE WITH FROM. MAILER MESSES IT UP
%this 
We see that there is an exponential decrease of
the coupling between the $i$th and $j$th skyrmions
for $R \gg 1/\kappa$.  
The XY term of Eq.(\ref{12}) favors forming a Q-skyrmion ($Q=N$) by 
recombining the $N$ single skyrmions, hence the collapse of the lattice.
(Note that the sign of this term in the antiferromagnet ordering is 
negative, therefore smaller seperation between skyrmions, $R$, is favorable.)
Such a combination costs the Coulomb energy. 
The global minimum solution of the energy functional can be identified
by the skyrmions self energy and the interaction terms.
Apart from the self energy terms which leads to an important effect  
to the recombination of the single skyrmions, one has to take the interaction
terms into account to predict the proper critical point of this transition.
Since the Coulomb interaction is long range 
and the XY term contributes to the total energy as a short range interaction,
then an antiferromagnet ordering within a square lattice can be the global  
minimum of the total energy, and no combining takes place except %SMG unless 
for small values of the Zeeman energy in agreement with
the recent result of Lillieh\"o\"ok {\em et al.} \cite{Lilliehook}.
We examine this effect by considering a special case, e.g. two skyrmions 
system. However, finding the global minimume of the total energy
for the general $N$-skyrmion system as a function of the filling factor
and the Zeeman splitting factor needs more investigation.
An estimate based on Eq.(\ref{12}) shows that 
the energy of a skyrmion with charge two is lower than the 
energy of two skyrmions each with charge one. Note that the inter skyrmionic 
XY and Coulomb energies are canceled %SMG
out if 
$g \sim |\nu-1|^{1/3}$ at the limit of $\nu \rightarrow 1$. 
Let's consider a system of two single skyrmions with charge one.
The total energy of this system can be obtained by employing Eq.(\ref{12})
into Eq.(\ref{8}). One may compare the total energy of the two single
skyrmions with the self energy of a skyrmion with charge two in order to
estimate the critical value of the Zeeman energy
\begin{equation}
E[{\bf m}; Q=2] = 2 E[{\bf m}; Q=1] + \frac{e^2}{2\epsilon_0}
\int_1 d{\bf r} \int_2 d{\bf r}' 
\frac{\rho_1({\bf r})\rho_2({\bf r}')}{|{\bf r}-{\bf r}'|}
-\frac{c^2\tilde{g}}{4\pi^2} K_0(\kappa|{\bf R}_1-{\bf R}_2|),
\label{new1}
\end{equation}
where $E[{\bf m}; Q]$ is the self energy of the classical Q-skyrmion. 
This can be evaluated by Eq.(\ref{eq1}) \cite{Lilliehook}. 
Results of our numerical calculation for $\tilde{g}_c$ for several
$\nu$ are listed in Table \ref{Tab1}. The values of the spin stiffness 
corresponding to the different fractional filling factors,
have been evaluated by Moon {\em et al.} \cite{Moon} numerically 
using the hypernetic chain approximation.
We expect that our expression for
the inter skyrmionic XY interaction, Eq.(\ref{new1}),
fails %SMG being failed 
for filling factors, $|\nu-1| > 4\pi(\kappa\ell_0)^2$ due to increasing the 
density of the skyrmions ($\kappa R \sim 1$). 
At $\nu=1$, the inter skyrmionic
interaction is exactly zero and one may find $\tilde{g}_c=0.53\times 10^{-5}$ 
in agreement with the Lillieh\"o\"ok {\em et al.} \cite{Lilliehook}.
At $\nu=1$, the competition between self energies yields
the transition, since the interaction terms do not play role in this
case. However, at $\nu < 1$ the value of $\tilde{g}_c$ is enhanced by the 
skyrmion interaction.
Since the order of the Coulomb energy and the short range XY interaction 
with respect to the number of particles are ${\cal O}(N^2)$ and 
${\cal O}(N)$ respectively, then the effective energy of a multi-skyrmion
system, Eq.(\ref{12}), becomes dominant by Coulomb energy,
hence the stability of the lattice at the limit of large N as 
a local minimum is guaranteed.
For larger separations, the XY term of Eq.(\ref{12}) is
therefore negligible compared to the effective Coulomb energy. 
This implies that for a lower density of skyrmions 
the Coulomb interaction determines the lattice
structure and leading to a triangular lattice.
%, hence, no recombination of skyrmions will occur.  %SMG
However, for a more dense system one may expect a 
different stable structural symmetry, in agreement with the calculation of 
Brey {\em et al.} \cite{Brey}.
One may define a domain of separation %SMG 
for which the %SMG that the 
square lattice is valid, e.g.
$\nu_{2c} < \nu < \nu_{1c}$ or $\kappa^{-1} < R < R_0$. 
Here the length scale 
cut off is denoted by $R_0$, beyond that a phase transition
to a triangular lattice occurs. 
%Note that the electrostatic dipole term
%may be significant at small Zeeman splitting
%where the overlap between different skyrmions becomes significant. 
%In this situation the effective interaction between skyrmions can 
%be described by the electrostatic dipole interaction then
%the square lattice may survive for a dense skyrmions
%even at zero Zeeman splitting. 
One may consider the static and dynamical properties of a skyrmion 
lattice \cite{Henk} using the interaction that we have derived 
\begin{equation}
V(|{\bf R}_i-{\bf R}_j|; \chi_i - \chi_j) =
  V_0(|{\bf R}_i-{\bf R}_j|)
     + \cos(\chi_i - \chi_j) V_1(|{\bf R}_i-{\bf R}_j|)~.
\label{hh1}
\end{equation}
In our model $V_0(R)$ is the electrostatic monopole interaction and 
$V_1(R) = \frac{c^2\tilde{g}}{4\pi^2} K_0(\kappa R)$. This is consistent 
with the result of variational calculation \cite{Henk,Shankar,Nazarov}.
Obviously, the skyrmion density may be controlled %SMG
via the Landau level filling factor.
As mentioned above, unlike the case where the 
competition between monopole Coulomb term and the 
gradient dipole terms is crucial %SMG
to the stability of the square lattice,
the electrostatic monopole interaction is the dominant %SMG ted 
term to determine the structure of the lattice for an 
extremely dilute ($\nu \sim 1$) skyrmions. In this situation a specific 
skyrmion moves in a background which %SMG
is being made by the vacuum of the other
skyrmions. Since $u_i$ which has been defined by Eq.(\ref{3}),
is zero every where but in the $i$th region,
then $\rho({\bf r})=\sum_i \rho_i({\bf r})$. 
Clearly the total charge is the summation upon the  
individual skyrmion's %SMG
charge, i.e. %SMG
$Q_{\rm tot}=\sum_i Q_i$ where $Q_i$ is the topological 
charge of localized skyrmion in that region. In this case,
the interaction between separated skyrmions is independent
of the relative orientation.
The skyrmions may be considered by point particles where
$\rho_i({\bf r})=\nu \delta({\bf r}-{\bf R}_i)$ and
\begin{equation}
V_{\rm eff} =
\frac{e^2}{2\epsilon_0} \sum_{i\neq j}^N
\int d{\bf r}\int d{\bf r}^\prime \;
\frac{\rho_i({\bf r})\rho_j({\bf r^\prime})}
{|{\bf r}-{\bf r}^\prime|} =
\frac{\nu^2 e^2}{2\epsilon_0} \sum_{i\neq j}^N
\frac{1}{|{\bf R}_i-{\bf R}_j|}.
\label{15}
\end{equation}
Therefore the ground state is being described by Eq.(\ref{15}) is clearly
a triangular lattice, independent of the skyrmion orientation, 
i.e., a  Wigner crystal \cite{Bonsall}. This is in agreement with the
result of Green {\em et al.} \cite{Green}.
%One may expect that the orientation of skyrmions at $\nu=1$ becomes aligned 
%to a specific direction due to spontaneous symmetry breaking
%of the ground state to reduce the Coulomb
%exchange energy of Eq.(\ref{15}) where at this filling factor 
%the skyrmions obey the Fermi statistics \cite{Nayak,KunYang}.
%This opens the possibility of the non-zero temperature Kosterlitz-Thouless
%phase transition at $\nu=1$ due to the formation of vortices from
%the skyrmions hedgehog fields.

One may find the critical filling factor $\nu_{1c}$ 
where a transition
into square lattice takes place. 
Note that we have made an approximation 
associated with $R > \kappa^{-1}$ to demonstrate the 
existence of the square lattice. 
We may expect the approximation, namely the 
superposition rule of Eq.(\ref{5}), fails 
for a high density skyrmions. The crossover between low 
and high density of skyrmions % SMG may 
occurs at $\kappa R \sim 1$
and leading to the lower critical filling factor $\nu_{2c}$, i.e.
$\nu_{2c} = 1 - \tilde{g}/2\rho_s$. Using the typical Zeeman energy, 
$\tilde{g} = 0.015 e^2/\epsilon_0\ell_0$ for GaAs and taking the
advantage of skyrmion-antiskyrmion duality, yields
$|\nu_{2c}-1| \sim 0.3$. 
Our estimate for lower limit of $\nu_{c}$ 
shows good agreement with the result of
random phase approximation of Brey {\em et al.} \cite{Brey}.

The appropriate quantum mechanical spectrum of the Skyrme lattice 
in our model can be obtained %SMG emerged 
through Eq.(\ref{12}) and the Wess-Zumino action
\begin{eqnarray}
&&\left\{ \sum_{i=1}^N 
\left[ -4\pi\rho_s\ell_0^2 (\nabla^2_{\bf R_i} + \kappa^2) +
\frac{1}{2}\Lambda_0^{-1}
\left(-i \frac{\partial}{\partial \chi_i} - \xi_0 \right)^2 
\right] + \sum_{i \neq j} V_0(|{\bf R}_i-{\bf R}_j|)\right.\nonumber\\&&
\left. + \sum_{<ij>} V_1(|{\bf R}_i-{\bf R}_j|)\cos(\chi_i-\chi_j) \right\} 
\Psi([{\bf R}; \chi]; t)
=i\hbar \frac{\partial}{\partial t}
\Psi([{\bf R}; \chi]; t),
\label{12.4a}
\end{eqnarray}
%It can be shown by Eq.(\ref{12.4a}) that
%there are two different goldestone modes \cite{Girvin} where
%the ground state has two broken symmetries.
%One is associated with U(1) spin rotation about the Zeeman field
%and the other is related to translational 
%symmetry breaking due to existence of a square lattice.
%The excitation states of the former are spin waves and the latter is
%2D phonons. 
where  $[{\bf R}; \chi] \equiv 
({\bf R}_1; \chi_1, {\bf R}_2; \chi_2, ..., {\bf R}_N; \chi_N)$
and ${\bf R}_i \equiv (R_i\cos\varphi_i, R_i\sin\varphi_i)$.
$\varphi_i$ is the standard azimuthal angle, indicating 
the $i$th skyrmion. $\Lambda_0=\xi_0/(3\tilde{g})$ is the moment of inertia
of a single skyrmion, and $\xi_0 = E_z/(2\tilde{g})$. 
Obviously, Eq.(\ref{12.4a}) describes a quantum rotor problem for 
$\chi$-degree of freedom \cite{Steve}, consistent with the result 
of Ref.\cite{Cote97,Nazarov}. This model has been used 
to describe the superconductor-insulator transition in granular
superconductors and Josephson junction arrays \cite{Cha}.
We may expect an insulator-superconductor phase transition occurs due to
variation of the Zeeman splitting ($\tilde{g}$) at a given filling factor
($\nu < 1$). 
Using the empirical %SMG
data for practical systems which are consistent with the 
experiments of Bayot {\rm et al.} \cite{bayot}
($\tilde{g}=0.015 e^2/\epsilon\ell_0 \sim 2 K$ at $\nu=0.8$)
in which the XY coupling constant, 
$J = (2\tilde{g}/\pi^2) K_0(\sqrt{\tilde{g}/2\pi\ell_0^2\rho_s} \; R) \sim 
0.05 K$, is a fraction of the $(2\Lambda_0)^{-1} \sim 0.4 K$
leads to the prediction of the Mott insulating phase for this system
where the position of skyrmions are fixed. 
We also expect that the system go through superconducting phase
for $J$ greater than $\Lambda_0^{-1}$ since the skyrmions carring
the electrical charges.
The ground state described by Eq.(\ref{12.4a}) 
for skyrmions orientation is a quantum antiferromagnet where 
the U(1) symmetry is broken spontaneously. 
Here $E_z = 0.214 e^2/\epsilon\ell_0$ is the Zeeman energy associated with
the optimal single skyrmion solution corresponding to 
$\tilde{g}=0.015 e^2/\epsilon\ell_0$, is evaluated by
the solving the non-linear, non-local integro-differential equation
Eq.(\ref{eq3}), for a single skyrmion numerically. 
The details of this calculation will
be presented elsewhere \cite{Abolfath}.

The Hamiltonian, Eq.(\ref{12.4a}), may also lead to a finite
temperature %SMG
Kosterlitz-Thouless phase transition which might be observable.
We may predict 
%two different 
a critical temperatures %SMG
for the Kosterlitz-Thouless phase transition based on the classical XY-model.
%and the quantum rotor model.
Our calculation for the classical 2D nearest neighbor XY-model 
on the square lattice, shows that the phase transition % SMG may 
occurs
around $T_c = 0.9 J \sim 0.04 K$ at $\nu=0.8 \; (B_0 \sim 7 T)$
where we consider the square lattice constant, $a=2\ell_0\sqrt{\pi/|\nu-1|}$,
and the nearest neighbour separation between skyrmions, $R=a\sqrt{2}/2$. 
Recent observations of an anomaly in the heat capacity %SMG
may support such a phase 
transition \cite{bayot,Allan96}. However, this interpretation
can be controversial since one may argue that this effect
is a remnant %SMG reminiscent 
of the nuclear Schottky anomaly \cite{Steve,Cote97}. 
%Our prediction based for the quantum rotor 
%model yields $T_c = \tilde{g} \sqrt{3J/E_z} \sim 0.14 K$ at the 
%same filling factor and 
%showing agreement with the numerical Hartree-Fock
%calculation of C${\rm \hat{o}}$te {\em et al.} \cite{Cote97}.

%%%%%%%%%%%%%%%%%%%%%%%%%%%%%%%%%%%%%%%%%%%%%%%%%%%%%%%%%%%%%%%%%%%%%%%%%%%

\section{Conclusion} 
We showed that for extremely dilute skyrmions, the ground state
is a triangular lattice to minimize the Coulomb repulsion. However
at higher densities, a square lattice forms to optimize the spin 
gradient and Zeeman energies. 
The ferromagnet triangular lattice become frustrated by
increasing the skyrmion density at critical filling factor
$\nu_c$ and a transition occurs into an antiferromagnet
square lattice. We also argued the possibility of the 
Kosterlitz-Thouless and superconducting-insulator phase transition.
We have shown that in some special cases the multi-skyrmionic
system becomes unstable with respect to the formation of the
single-skyrmion system with the same conserved topological charge.

\section{Acknowledgement}
The authors thanks H.A. Fertig, S.M. Girvin, A.H. MacDonald, J.J. Palacios, 
S. Rouhani and H.T.C. Stoof for helpful discussion.
Our special thanks go to S.M. Girvin and H.T.C. Stoof for %SMG the greatful
inspiring comments and ideas on the subject.
The work at Indiana University is supported by NSF DMR-9416906.
MA would like to thank Indiana University for the warm hospitality.

\begin{table}
\caption{
The critical $\tilde{g}$-factor for several Landau level filling factors
are presented. The two single-skyrmions system become unstable at
$\tilde{g} \leq \tilde{g}_c$ and a transition to a single skyrmion system
with charge %SMG
two occurs.
}
\begin{tabular}{lcc}
$\nu$ & $\rho_s (e^2/\epsilon\ell_0)$ & $\tilde{g}_c (e^2/\epsilon\ell_0)$ 
\\ \hline
1.0 & 2.49 $\times 10^{-2}$ & 0.53 $\times 10^{-5}$ \\
1/3 & 9.23 $\times 10^{-4}$ & 0.19 $\times 10^{-3}$ \\
1/5 & 2.34 $\times 10^{-4}$ & 0.52 $\times 10^{-4}$ \\
\end{tabular}
\label{Tab1}
\end{table}

\end{document}